\documentstyle[12pt]{article}

\begin{document}
\begin{titlepage}
\begin{center}
\large\bf On the Formation of Transient (Na$_{19}$)$_2$ and (Na$_{20}$)$_2$
Cluster Dimers from Molecular Dynamics Simulations\\
\normalsize\bf
\vspace{1.0cm}
F.S. Zhang$^{1}$, F. Spiegelmann$^{2}$, E. Suraud$^{2}$, V. Frayss\'e$^{1}$,\\
\vspace{0.3cm}
R. Poteau$^{2}$, R. Glowinski$^{1}$ and F. Chatelin$^{1}$\\
\vspace{0.5cm}
\normalsize

{\it
$^{1}$GIP, CERFACS, 42 Avenue G. Coriolis, 31057 Toulouse Cedex, France\\
\vspace{0.3cm}
$^{2}$Laboratoire de Physique Quantique, Universit\'e Paul Sabatier,
      118 route de Narbonne, 31062 Toulouse Cedex, France\\ }
\end{center}

\vspace{0.5cm}
\centerline{ {\bf Abstract} }
\vspace{0.1cm}
\begin{quote}

By using tight binding molecular dynamics simulations, we discuss
the possibilities to form (Na$_{19}$)$_2$ and (Na$_{20}$)$_2$ cluster
dimers in sodium cluster collisions. In the case of Na$_{19}$ + Na$_{19}$,
we show that the formation of a prolate dimer-like (Na$_{19}$)$_2$ maybe
depend on the initial relative orientations of the
colliding clusters. A similar
study for Na$_{20}$ + Na$_{20}$ does not seem to show the same dependence on
the initial orientations in the formation of the (Na$_{20}$)$_2$ cluster dimer.

\vspace{1.0cm}
\end{quote}
\end{titlepage}

\begin{doublespacing}
In the very recent years, there has been several attempts to study
cluster cluster collisions of alkali metals. These studies tend to
investigate parallel behaviours between nuclear physics and cluster physics.
Indeed, as in the case of nuclei, it was found that the particular
stabilities of some alkali metal clusters\ \cite{jelexp}
correspond to closed shell electronic structures of the spherical
jellium model \ \cite{jel1} or its spheroidal and ellipsoidal
extensions\ \cite{jel2}. However there are  additional  peaks in
mass spectra which do not correspond to shell closure, such as for instance
Na$_{38}$\ \cite{exp}. Obviously, one should be careful about the real
meaning of the abundancies in mass spectra, first because some of the
observed peaks are significantly dependant upon the
experimental conditions, and second because those abundancies may have both
an energetic and  a kinetic origin.
Nevertheless special attention has been paid to the Na$_{38}$ case which was
found to present an abundancy in mass spectra almost as large as Na$_{40}$.
By using a two-center jellium background,
Saito and Ohnishi\ \cite{saito1} proposed the concept of  a
(Na$_{19}$)$_2$ cluster dimer as an intermediate step in the process of
formation of a compact structure Na$_{38}$. Indeed, they found that the dimer
binding energy $E_b(D_{CM})=E_p(D_{CM})-E_p(\infty)$
where D$_{CM}$ is the intercluster  center of mass (CM) separation,
has an attractive local minimum at 16.5 Bohr, which is separated by a  barrier
from the lowest energy  minimum corresponding to a united cluster. In their
work, the barrier was found to be  0.5 eV {\em above} the asymptote. The same
analysis has been applied to (Na$_4$)$_2$ and (Na$_8$)$_2$ dimers by
Ishii, Saito and Ohnishi\ \cite{saito2} and to the dimer decay of
potassium clusters by Saito and Cohen\ \cite{saito3}. Furthermore,
Nakamura {\em et al.}\ \cite{saito4} used a similar approach for the
discussion of cluster fragmentation. More recently, Engel
{\em et al.}\ \cite{schmidt3} carried out a two-center jellium background
calculation for Na$_{38}$ but they found that the barrier between the
long distance well and the compact cluster
potential well lied 0.27 eV {\em below} the
asymptotic dissociation. From a dynamical point of view, this last result
is significantly different from the one of Saito and Onishi, since it means
that the collisional system always has enough energy  (even from a classical
dynamics point of view) to overpass the barrier towards the compact zone.

   It should be noted that it is not completely
obvious that the picture of a two center static
homogenous jellium model, with only one explicit reaction coordinate (namely
the
intercluster distance) provides a realistic account of an actual reaction
path with 3(2n)-6 degrees of freedom for the collision of two Na$_n$ identical
moities. In particular, local reorganisation (taking into account other
degrees of freedom than the intercluster separation) of the ions during
the collision might lower the barrier between the
dimer-like and compact structures.
Beside the static aspect of the potential energy surface, especially
if the barrier lies below the dissociation limit, the main
problem is to determine the lifetime of the transient dimer-like species, which
of course will depend on the
specific shape and height of the  separation barrier  along the reaction
path, but may also depend on the kinetics of the reaction.  Schmidt
{\em et al.}\ \cite{schmidt1,schmidt2} presented molecular dynamics
simulations for collisions of smaller systems, namely
Na$_8$ + Na$_8$ and Na$_9$ + Na$_9$. They reported that a (Na$_{9}$)$_2$
dimer-like structure exists about 7200 fs (1 fs = $1\times 10^{-15}$ s) after
contact and they showed a prolate structure  obtained by optimizing
the system configuration at 960 fs (close to the beginning of the simulation).
After about 7200 fs, the (Na$_{9}$)$_2$ dimer-like was reported to decay
towards a compact Na$_{38}$ with significant thermal excitation.

In this paper, we present Distance Dependent Tight Binding Molecular
Dynamics (DDTB-MD) simulations for Na$_{19}$ + Na$_{19}$ and
Na$_{20}$ + Na$_{20}$ with two specific initial relative orientations of the
colliding clusters at a CM incident energy
per atom of $E_{CM}/N=0.0125 eV$. This
total CM collision energy is in magnitude less than 0.2 times of the
dissociation energy of Na$_{38}$ or Na$_{40}$ at equilibrium
into identical moieties.
For the Na$_{19}$ + Na$_{19}$ system, we find different results,
depending on the initial orientation. In one case, a transient dimer-like
structure (Na$_{19}$)$_2$ is found to exist for about 3000 fs,
while in the other case this structure does not live a significant
amount of time (some hundreds of fs at most). On the contrary, for
the Na$_{20}$ + Na$_{20}$ system, a transient dimer-like structure
(Na$_{20}$)$_2$ is found whatever the initial orientations are.

The model used here for representing the ground state Potential Energy
Surface (PES) of the total system is
a Distance Dependent Tight Binding hamiltonian.
The energy function for a given configuration of
the nuclei is computed according
to the monoelectronic hamiltonian\ \cite{ups}
\begin{eqnarray}
\hat{h} &=& \sum_{i,j}h_{ij}a_i^+   a_j
\nonumber \\
 h_{ii} &=& h_{ii}^{(0)} + h_{ii}^{(2)}
\nonumber \\
        &=&  \sum_{k\neq i} \{ \rho_{ss}(R_{ik})
           - \frac{t^2_{s \sigma} (R_{ik}) } {\epsilon_{3p} - \epsilon_{3s} }
\}
\nonumber \\
 h_{ij} &=& h_{ij}^{(0)} + h_{ij}^{(2)}
\nonumber \\
        &=& t_{ss} (R_{ij}) - \sum_{k\neq i,j} \{
          \frac{t_{s \sigma} (R_{ik}) t_{s \sigma}
             (R_{jk}) } {\epsilon_{3p}-\epsilon_{3s}}
          \times \frac{ {\bf R_{ik}} \cdot {\bf R_{jk}}               }
	  	        { \mid {\bf R_{ik}} \mid \mid {\bf R_{jk}} \mid } \}
\end{eqnarray}
where $a_i^+$ and $a_j$ are creation and annihilation operators
corresponding to $s$ orbitals on site $i$ and $j$, repectively.
The matrix elements $h_{ij}$ which are distance-dependent include
perturbatively the effect of $p$ orbitals. The three functions
$\rho_{ss} (R),t_{ss} (R),$ and $t_{s \sigma} (R)$ represent
the ion-ion repulsion, the $s-s$, and the $s-p_{\sigma }$ transfer
integrals at interatomic distance $R$, respectively. This parametrization
was adopted in order to reproduce almost exact
$^1 \Sigma _g^+$ and $^3 \Sigma _u^+ $ potential curves of Na$_2$
and the dissociation energy of Na$_4$ as obtained in $ab$ $initio$
Configuration Interaction calculations. The total electronic
energy in the Born-Oppenheimer approximation is computed as
the sum of the one-electron eigenvalues
\begin{equation}
E=\sum_{i \in occ} n_i \epsilon _i \qquad .
\label{Etot}
\end{equation}
This Tight-Binding was shown, in a previous work \cite{ups}, to
provide geometrical structures and stabilities in good agreement
with more sophisticated calculations and also
with experiments \cite{toappear}.

The derivatives of the total energy according to the Cartesian
coordinates are calculated by the Hellmann-Feynman theorem.
We employ the DDTB\ \cite{ups} and Monte-Carlo simulated-annealing
technique\ \cite{mcsa} for determining the stable configurations
of projectile and target clusters. We initially prepared the colliding
clusters with two different particular relative orientations which
are illustrated in Fig. 1. In the first case,
herafter refered to as (A), the principal axes of inertia of
the two clusters are aligned. In the second case, one of the cluster has been
rotated by a 90 degree angle around an axis perpendicular to the collision
direction. This case is herafter refered to as (R).
The collective rotational modes were discarded. Both
projectile and target clusters were prepared with a very low temperature(18K).
Newton's equations of motion are integrated numerically,
\begin{equation}
m_i {\bf \"r}_i = -\frac{\partial E} {\partial {\bf r}_i}
\end{equation}
where $E$ is the Born-Oppenheimer (BO) surface (Eq. (\ref{Etot})),
$m_i$ and ${\bf r}_i$ are
the mass and coordinate of i-$th$ atom. Time step of 1 fs have proven to
guarantee the conservation of energy in the present work.
The vibration of the separated clusters was allowed to reach thermal
equilibrium by propagating the internal coordinates during 2000 fs before
time t = 0  which corresponds  to the instant at which the translational
motion is switched on. The averaged potential energy during this
thermalization phase are -27.642 eV for Na$_{19}$ + Na$_{19}$ and -29.595 eV
for
Na$_{20}$ + Na$_{20}$. The initial intercluster separation is big enough
(60 Bohr) to ensure that there is no interaction between projectile and target.

In Fig. 2, we show the time evolution of the intercluster separation
D$_{CM}$ (Bohr), and the axial ratios X$_0$, Y$_0$ and Z$_0$
for central collisions of Na$_{19}$ + Na$_{19}$ in cases (A) and (R).
The axial ratios X$_0$, Y$_0$ and Z$_0$ correspond to the analysis of
cluster structure in terms of an ellipsoid which can be calculated from
the principal moment of inertia I$_x$, I$_y$ and I$_z$ as
    X$_0 = \sqrt { \frac {I_x} {I_x + I_y + I_z} }$,
    Y$_0 = \sqrt { \frac {I_y} {I_x + I_y + I_z} }$,
and Z$_0 = \sqrt { \frac {I_z} {I_x + I_y + I_z} }$.
Both the intercluster separation and the axial ratios can give us
global geometric features of the system during
the collision process. From the simulations, the
collision process maybe essentially divided
into four stages: preparation, approach,
transient cluster molecule formation, and compound compact structure formation.
In the approach phase, the potential energy of the global system begins to
decrease as soon as  the attractive forces between the two clusters become
non-vanishing (D$_{CM}$ $\leq$ 25 Bohr), while the kinetic energy is
accordingly
increased. At about 3500 fs, the two clusters touch each other. It is clear
that
in case (A), the intercluster separation almost reaches its final equilibrium
values (D$_{CM}$ = 6 $\sim$ 8 Bohr) in about only some hundred fs after the
approach process (Fig. 2(a) solid line). In contrast, in case (R), the reaching
of the compact stage is delayed up to 6500 fs (Fig. 2(a) dashed line). An
apparent
difference also occurs for the axial ratios. In Fig. 2(b) the axial
ratios X$_0$, Y$_0$ and Z$_0$ almost
immediately reach a common value around 1.0 which
is typical of a spherical-like structure with some thermodynamical fluctuations
after about 4000 fs. For case (R) in Fig. 2(c), the axial ratio
Z$_0$ appears to reach 1.0 with a stepwise decay only after about 6500 fs. This
means that a transient prolate (Na$_{19}$)$_2$ cluster dimer can live
about 3000 fs in our simulations. This is partly consistent with the previous
findings
of Schmidt {\em et al.}\ \cite{schmidt1,schmidt2} for (Na$_9$)$_2$.
However, such a dynamical metastability only occured in case (R)
and has almost vanished in case (A). Thus, the lifetime seems to be strongly
influenced by the initial conditions.

As concerns the energetics, the time evolution of the potential energy
and the evolution of  the binding energy as a function of the intercluster
distance along the trajectory are given in Fig. 3 for cases (A)
and (R) respectively.
The lowering of the potential energy following
immediately the approach phase is slightly  different for the two cases. This
phase is essentially dependant on the geometry.
After contact, the potential energy is immediately lowered down to -28.5 eV in
case (A) (see Fig. 3(a)), while in case (R) (Fig. 3(b)), two
plateaus may be tentatively identified (a word of caution
is necessary here since this shift is partly smeared by the
fluctuations). The first plateau corresponds to
-28.33 eV in the stage between 3500 fs and 6500 fs whereas the second one
corresponds to -28.5 eV. In the variation
of the binding energy along the collision path, one may also see the
difference. The  potential energy undergoes a minimum
of about -0.71 eV at 15.5 Bohr, in case (R)
(Fig. 3(b)). This minimum is slightly deeper (-1.0 eV at 13.5 Bohr) in case
(A) (Fig. 3(d)). A more significant feature occurs at shorter distance
(between 9.5 Bohr and 12 Bohr) since the potential energy
corresponding to case (R) in this
range shows numerous oscillations. This provides further
evidence that the system keeps on fluctuating for some time
with a relatively prolate global geometry  and
an approximatively constant large intercluster distance. This  fluctuation is
almost absent for case (A). Thus the evolution
of the potential energy is consistent with the previous
discussion on the purely geometrical aspects.

We have also plotted in Fig. 3(b) and Fig. 3(d) the binding energy obtained
in the two-center jellium model (using the VWN results of
Engel {\em et al.}\ \cite{schmidt3}, dashed line). It is clear that
in the present simulations, if there exists a barrier between the prolate and
compact structures, it should be lower than the one reported by
the two-center jellium one-dimension model. This certainly
comes from the consideration of the 3(2n)-6 degrees of freedom in
the present work, in which numerous paths can be visited on the
potential energy surface, allowing for numerous local deformations
during the collision process. Thus the transient species can be understood
as a dynamical fluctuation between different isomers compatible with a
global prolate shape which has a finite lifetime. After a while, those
fluctuations drive the cluster towards a compact vibrationally
excited  structure.
In Fig. 4, we give some typical geometries for central collisions of
Na$_{19}$ + Na$_{19}$ at different times : 4000, 5000, 6000 and 15000 fs
in the (R) case. One may
notice from Fig. 4 that the mixing of the incident systems is only very partial
after 15000 fs, which means that the compact quasi-spherical cluster has not
become fully liquid-like (this is why in our simulation D$_{CM}$
is never reduced
down to zero as in the jellium case in which the distinction between the
initial
subsystems is meaningless in the final stage).

In order to check whether the Na$_{19}$ + Na$_{19}$ collision is a special
case for the formation of  a dimer-like cluster with reference to Na$_{38}$, we
have made the same simulations for Na$_{20}$ + Na$_{20}$. In both approaches
(A) and (R) we observed basically the same collision features as for
Na$_{19}$ + Na$_{19}$ in case (R), that is a transient
(Na$_{20}$)$_2$  species with prolate shape exists during a lifetime of about
3000 fs. So, the appearence of such a transient molecular cluster dimer
does not seem to be specific of the Na$_{38}$ situation(in Fig. 5 we only
show the results corresponding to case (R), those for case (A) are almost
identical).

In conclusion, from the Tight Binding Molecular Dynamics simulations in
central cluster-cluster collisions, we find that, in the case of large
Na$_{19}$ + Na$_{19}$ and Na$_{20}$ + Na$_{20}$ cluster collisions,
a transient dimer-like structure may exist during a few thousands fs.
However, this existence seems somwhat dependant on the initial orientions
of the colliding clusters. As already mentioned by Schmidt
{\em et al.}\ \cite{schmidt1,schmidt2} non zero impact parameters
might enhance the lifetime of such structures. In any case, the system
always evolves towards a compact thermally excited cluster.
{}From the present work, it is
not obvious that the dimer-like to compact transition
occurs readily between two well defined microscopic structures, but rather as
the result of complex fluctuations towards the final state. No
clear potential energy barrier such as described in the jellium model
seems to separate two specific regions. In
any case, further simulations are needed.
A real estimation of a dimer-like compound lifetime requires averages  over
numerous different initial conditions, such as orientations  and impact
parameters.
Work is in progress in order to adapt the code to parallel architecture
computers and achieve a more thorough study.

\end{doublespacing}

\newpage

\section*{Figure Captions}
\begin{description}

  \item{\bf FIG.\ 1.}\ \ \ Illustration for Na$_{19}$ + Na$_{19}$
   of the initially prepared colliding
   systems with two different particular relative orientations. In the first
case,
   refered as (A), the principal axes of inertia of the two clusters are
   aligned. In the second case (R), one of the cluster has been rotated by a 90
degree
   angle around an axis perpendicular to the collision direction.

  \item{\bf FIG.\ 2.}\ \ \ Time evolution of the intercluster separation
  D$_{CM}$ (Bohr) (a), and the axial ratios X$_0$, Y$_0$ and Z$_0$ for central
  collisions of Na$_{19}$ + Na$_{19}$ at incident energy per atom in CM system
  E$_{CM}$/N=0.0125 eV in cases (A) (b) and (R) (c), respectively.

  \item{\bf FIG.\ 3.}\ \ \ Time evolution of the total energy E$_t$, potential
  energy E$_p$ and the binding energy $E_b$ as as a function of intercluster
  separation D$_{CM}$ (Bohr) for central collisions of Na$_{19}$ + Na$_{19}$
  at incident energy per atom in c.m. system E$_{CM}$/N=0.0125 eV in cases
  of (A) ((a) and (b)) and (R) ((c) and (d)), respectively.

  \item{\bf FIG.\ 4.}\ \ \ Geometries for central collisions
  of Na$_{19}$ + Na$_{19}$
  at different time, namely, t= 4000, 5000, 6000 and 15000 fs,
  at incident energy per
  atom in c.m. system E$_{CM}$/N=0.0125 eV in case (R).

  \item{\bf FIG.\ 5.}\ \ \ Evolution of the intercluster separation D$_{CM}$
  (Bohr) (a), the axial ratios X$_0$, Y$_0$ and Z$_0$ (b), total energy E$_t$,
  potential energy E$_p$ (c), and the binding energy $E_b$ as as a function
  of intercluster separation D$_{CM}$ (Bohr) (d) for central collisions
  of Na$_{20}$ + Na$_{20}$ at incident energy per atom in CM system
  E$_{CM}$/N=0.0125 eV in case (R).

\end{description}
\end{document}